\documentstyle[prl,aps,preprint,epsf]{revtex}

\begin{document}

\draft

\tighten

\title{Stripe fractionalization I: the generation of Ising local symmetry}

\author {Z. Nussinov and J. Zaanen}

\address{Instituut Lorentz voor de theoretische natuurkunde, 
Universiteit Leiden, P.O. Box 9506, NL-2300 RA Leiden, The Netherlands}

\date{\today}

\maketitle

\begin{abstract}

This is part one in a series of two papers dedicated
to the notion that the destruction of the topological order
associated with stripe phases is about the simplest  
theory controlled by local symmetry: Ising gauge theory. This
first part is intended to be a tutorial- we will exploit the simple
physics of the stripes to vividly display the mathematical beauty of the gauge
theory. Stripes, as they occur in the cuprates, 
are clearly `topological' in the sense that the lines of charges are at
the same time domain walls in the antiferromagnet. Imagine that the
stripes quantum melt so that all what seems to be around is a singlet
superconductor. What if this domain wall-ness is still around in
a delocalized form? This turns out to be exactly the kind of `matter' 
which is described by the Ising gauge theory. The highlight of the
theory is the confinement phenomenon, meaning that when the domain wall-ness
gives up it will do so in a meat-and-potato phase transition. We suggest
that this transition  might
be the one responsible for the quantum criticality in the cuprates. In
part two\cite{strifractII} we will become  more practical, 
arguing that another phase is
possible according to the theory. 
It might be that this quantum spin-nematic has already been observed 
in strongly underdoped  $La_{2-x}Sr_x Cu O_4$.      
\end{abstract}

\narrowtext

\section{Dynamical generation of local symmetry.}

It has been a dream for a long time that
the profoundness of non-perturbative gauge theory could come alive
in the earthly forms of matter which are of interest to condensed
matter physicists. This is far from self-evident. Gauge theory
is controlled by local symmetry. The high energy physicist will
argue that it better be fundamental because local symmetry is infinitely
vulnerable towards explicit symmetry breaking influences making it global --
we refer to part II\cite{strifractII} for an example. However,
gauge theory carries a myriad of meanings and one usually exploits the 
principle that local symmetry implies a local conservation law
which in turn corresponds with a local constraint. This is at the
heart of the slave theories of electron fractionalization: one can
pretend that the electron (or Cooper pair) is actually a composite
particle, paying the price that the pieces of the electrons are 
minimally coupled to strongly interacting gauge fields. Although
theoretically consistent, the drawback is that these gauge fields
are highly mathematical entities lacking a material interpretation.
How to measure the $SU(2)$ gauge field of Lee and coworkers\cite{leegauge}? 
The $Z_2$ theory of Cooper pair fractionalization 
by Senthil and Fisher\cite{senthill}
is doing better in this regard but it leaves one wondering why the
vortices in the dual condensate should form pairs -- that is their
physics.

We\cite{philmag}, and others\cite{sachdev1,sachdev2}, 
were astonished when it became clear that
a feet-on-the-ground physics problem turned out to be in correspondence
with the most elementary gauge theory: the Ising- or $Z_2$ gauge
theory, invented by Wegner in the early 1970's\cite{wegner}, 
which has played an
important role in the early history of non-perturbative
gauge theories\cite{kogut,fradkin}. 
The physics problem is inspired by the strong empirical
case, as presented by e.g. Tranquada in this volume\cite{tranquada}, 
that superconductors of the $La_{2-x}Sr_xCuO_4$ variant 
have to do at the same time 
with a quantum disordered stripe phase. What can be said in general about 
such a stripe quantum liquid? `General' means here Landau's method: we 
are 
only allowed to use symmetry. Stripe order means that symmetries are broken,
and quantum liquid means that these symmetries are restored. The key
is that stripes break a variety of symmetries and some of these symmetries
might remain broken even in the liquid. States carrying this `partial order' 
might be easily
overlooked by the unprepared experimentalist. Typical examples are the
quantum smectic- and nematic states introduced by Kivelson, Fradkin, and 
Emery\cite{kivelson}. 

The states which will be discussed here are even more radical.
Empirically, the charge stripes are at the same time domain walls in the
spin system and in section II we will explain that this domain wall-ness
can be viewed as a form of long range order, albeit of a {\em geometrical}
nature. It follows that, in principle, this geometrical order can persist
when charge and spin become quantum disordered, such that the system
is, in the first instance. a singlet
superconductor. Such a superconductor carries a truly {\em topological}
order, and the mathematical description of this order is the essence
of the gauge theory (section III). Eventually this order also gets
destroyed in a normal phase transition: the meaning of confinement in 
the Ising gauge theory. However, it is impossible to directly observe the
topological order and its destruction by existing experimental machines 
(section IV).

\section{Our space and the space of the spins.}

What is stripe order? It consists, at the least, of three forms of
spontaneous symmetry breaking: (i)
the electrical charge breaks space translational and -rotational symmetry.
It is just a Wigner crystal with an interesting 
(orthorombic)  crystal structure.
Assuming that stripes are made from bosons (electron pairs), one can argue
on general grounds that upon the restoration of translational symmetry
superconducting condensation will follow automatically, although this state
might still break rotational invariance (the quantum nematic [10]). In
the remainder we will take this superconductivity for granted. (ii) Stripes
might also be co-linear antiferromagnets, breaking spin-rotational
symmetry. In a weakly coupled BCS superconductor the spin system is 
characterized by a quantum disordered singlet ground state,
but it seems now generally appreciated that this is a specialty of 
weak coupling.
Being different symmetries, nothing argues against a coexistence state where
the same electrons condense at the same time 
in a superconductor and an anti-ferromagnet.

Subsequently, this anti-ferromagnet can disorder in a spin-only quantum 
phase transition. This spin disordered state is a spin singlet and such
a vacuum is indistinguishable from a BCS superconductor, except than that
it can be characterized by very strong antiferromagnetic spin correlations.
This generalized view on superconductivity seems to be much closer to the
physics of the cuprate superconductors than the conventional weak coupling
view, see Sachdev {\em et  al.}\cite{sachdev1}. 
(iii) However, stripe order is more than just
spin and charge order. It is also order associated with the fact that 
the charge stripes are domain walls, $\pi$-phase boundaries
in the stripe anti-ferromagnet. The nature of this  unconventional
order is
the subject of this paper.

Empirically, this `domain wall-ness'
is clearly a form of order. Long range order means that one can predict what 
happens at $+\infty$
when one knows what is going on at $-\infty$. Taking two points infinitely
far apart in a stripe phase, one can count if there are an even or odd
number of charge stripes in between, and one can predict if the staggered
order parameters in the two  domains are parallel- or antiparallel relative
to each other. The key observation  is that this order can
persist in principle when spin- and charge are disordered,
so that the system is in any other regard a normal superconductor.

One has to ask first the simple question: 
we observe stripes to be domain walls, but domain walls {\em in what}? 
Domain walls have to
do with scalar fields (Ising systems) and surely neither the spin-
nor the charge systems have anything to do with Ising fields.       
Domain walls are topological entities and the topological excitations
associated which charge and spin  are different
(dislocations/disclinations and skyrmions, respectively).  The
proper answer is instead, {\em stripes are domain walls in the  
geometrical quantity sublattice parity}. It is a requirement for the
existence of colinear spin-order that the spin system lives in a
bipartite space, meaning that the lattice can be subdivided into an
$A$ and $B$ sublattice, such that every site on the A sublattice is     
surrounded by B sublattice sites and vice versa. This can be done
in two ways, $\cdots - A - B - A - B - \cdots$ or $\cdots - 
B - A - B - A - \cdots$. This $Z_2$ valued quantity we call the
sublattice parity $p = \pm 1$. Stripe `domain wall-ness' order means
that every time one passes a charge stripe $p$ changes sign: the
sublattice parity is ordered.

\begin{figure}[t]
\hspace{0.00 \hsize}
\epsfxsize=0.8\hsize
\epsffile{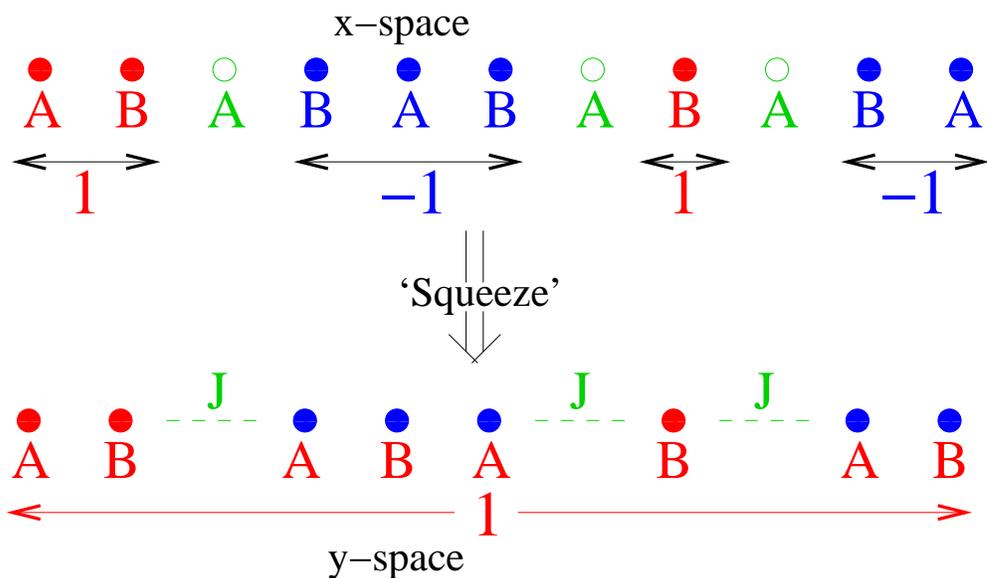}
\caption{The Ogata-Shiba mechanism for spin-charge separation in
the Luttinger liquid [13,14]. Distribute first the electrons over the
Hubbard chain and the amplitude of such a charge configuration
will only depend on the positions of the electrons (upper panel).
Take such a charge configuration, and take out the sites where the
holes are, and substitute these with exchange bonds (J) between the
spins of the electrons neighboring the holes (lower panel). The
spin system lives in this squeezed space. Relative to the real
sublattice division of squeezed space, sublattice parity in full
space flips every time a hole is passed.}
\label{fig1}
\end{figure}

How do we know? A proof of principle can be delivered in the one
dimensional context. As we demonstrated recently\cite{kruis}, the Luttinger
liquid derived from the 1D Hubbard model turns out to carry also a form of
topological sublattice parity order. This rests on the Ogata-Shiba
mechanism\cite{ogata} 
for charge-spin separation as deduced from the Bethe-Ansatz
solution in the large $U$ limit which we demonstrate to be universal
in the scaling limit for all positive $U$'s. The remarkable observation
by Ogata-Shiba is that the structure of the Bethe-Ansatz wave function
is explicitly of a geometric nature, as illustrated in Fig. 1.
In a first step, distribute the electrons over the chain and such a
charge configuration will acquire an amplitude equal to that of a
system of spinless fermions. This configuration of electron charges
in turn defines a {\em new space} in which the spin system lives,
obtained by removing the sites where the holes are, substituting 
these missing sites by antiferromagnetic exchange bonds between the
spins neighboring the hole: the `squeezed space'. Hence, spin-charge
separation means that the spin dynamics is just governed by a spin-only
Heisenberg problem defined in squeezed space. However, this spin dynamics
cannot be observed directly by doing measurements in the full chain.
In the unsqueezing operation, every time one inserts a hole one has also
to add a site and this in turn means that the `true' spin dynamics of
squeezed space gets modulated in addition with a flip in the sublattice 
parity attached to the hole: see Fig. 1.

What is the relationship with the 2+1 dimensional case? Just take the
Ogata-Shiba mechanism as principle, to see how it should be
implemented  in higher dimensions.
A requirement is that the lattice in the higher dimensional space is 
bipartite, which is fortunately the case for the square lattice of the
cuprates. Insist that the spin system lives on this unfrustrated square
lattice. Reinsert the charges and unsqueeze this lattice: 
this can only be accomplished when the
charges form lines which are uninterrupted domain walls in the sublattice
parity. In other words, stripes are at least symmetry-wise quite
like the 1+1D Luttinger liquids. This should be appreciated as a mere 
phenomenological argument: given that long-range `domain wall-ness' order
is observed, it has to be that squeezed space exists and that the 
domain walls are carried by sublattice parity. The microscopic origin 
of the domain wall-ness
is a different matter. A fair understanding exists where it is coming
from and we refer to the stripe literature\cite{zastripes}: it is about
avoiding  the frustrations caused by the quantum motions of a single hole
in the antiferromagnet by organizing the holes in stripes. At the same
time, it does not have to happen in 2+1D.
Hence, the domain wall-ness   can be destroyed in favor 
of a conventional, Fermi-liquid/BCS state and this
is the subject of the remainder of this paper.

\begin{figure}[t]
\hspace{0.00 \hsize}
\epsfxsize=0.69\hsize
\epsffile{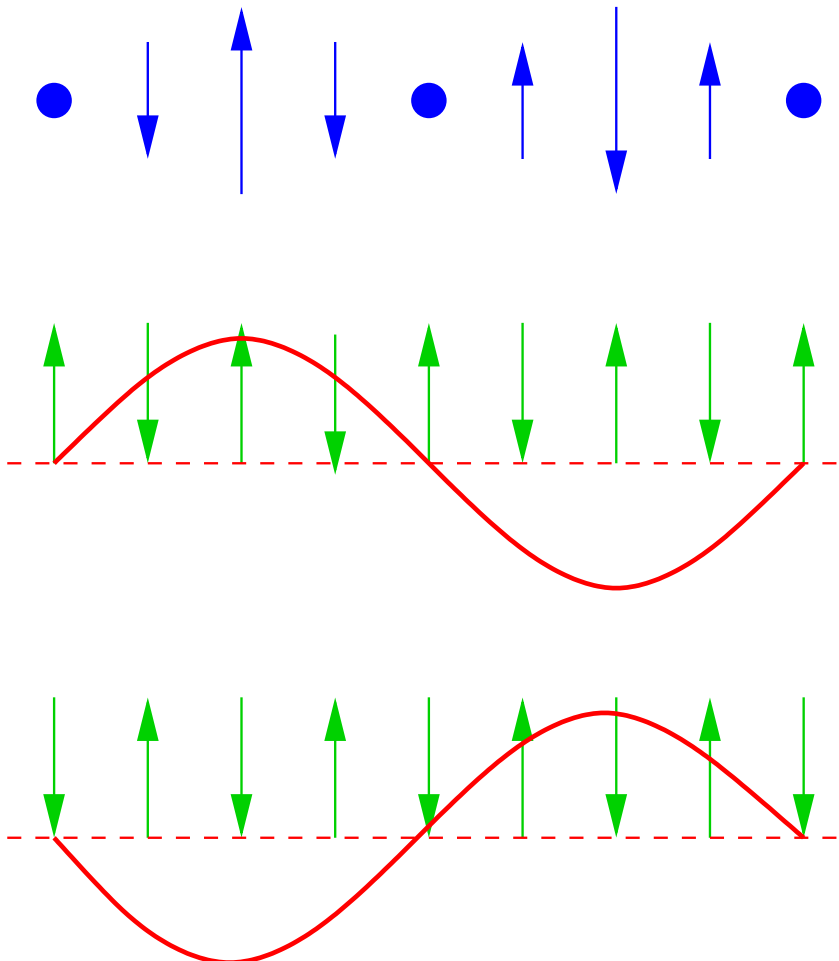}
\caption{Insofar as basic symmetries are concerned, spin density waves are
just the same as stripes. Implicit in the description of the spin density
wave order parameter is a $Z_2$ gauge redundancy [5,6]. 
The spin density wave
order parameter (upper panel) is a product of  staggered spin and amplitude,
and the shift in sublattice parity in the former can always be compensated
by shifting the phase of the amplitude envelope by $\pi$.}
\label{fig2}\end{figure}

On a side, we notice that with regard to symmetry stripes
cannot be distinguished from incommensurate co-linear spin density
waves. Let us consider the sublattice parity order in this context,
using an argument due to Sachdev\cite{sachdev1,sachdev2}.  
The order parameter 
is $O_{SDW} (\vec{r})= \Phi (\vec{R}) \vec{M}(\vec{r})$;
$\vec{M} (\vec{r}) = \langle (-1)^{\vec{r}} \vec{S} \rangle$ is 
the expectation value for staggered spin, while $\Phi (\vec{r}) = 
\cos( \vec{q} \cdot \vec{r}) \Phi$ is the {\em amplitude} of the spin density
wave. Implicit to this description is a $Z_2$ gauge redundancy: $
\Phi \rightarrow - \Phi, \vec{M} \rightarrow - \vec{M}$ leaves the 
physical $O_{SDW}$ unchanged, see fig.3. 
For a static spin density wave this does 
not carry any consequence. However, spin amplitude and charge are
fundamentally indistinguishable and the SDW should be considered as
a special, weakly coupled limit of the static stripes. When charge is 
disordered, spin amplitude is disordered as well. Assuming that staggered
spin $\vec{M}$ stays ordered the $Z_2$ gauge becomes alive: $\vec{M}
\leftrightarrow -\vec{M}$ which is the same as the sublattice parity
gauge transformations $\cdots - A - B - \cdots \leftrightarrow
 \cdots - B - A - \cdots$. The state characterized by
$\langle \vec{M} \rangle \equiv - \langle \vec{M} \rangle$ is the
spin nematic which is discussed in paper II.

\section{Geometry, topology and Ising gauge theory}

Gauge theory has interesting connections with geometry and topology
and a vivid, simple example follows from the stripe interpretation of
$Z_2$ gauge theory: it is at least a tutorial tool to explain
to students some of the basics.

Obviously, static stripes have noting to do with Ising local symmetry.
This comes alive only when the stripes are truly delocalized. The key is
that the topological excitations associated with the destruction
of the charge order and the sublattice parity order are different,
with the consequence that charge disordered (superconducting) states
can exist carrying a truly topological (non-local) domain wall order.
Ising gauge theory is the theory describing this  topological order, and its 
destruction.

This is easy to see: the restoration of translation symmetry is uniquely
associated with the spontaneous proliferation of dislocations, Fig. 3.
When the crystal is bosonic (e.g., build from Cooper pairs) a theorem
due to Feynman insists that nothing can prevent the system to become
a superconductor when translational symmetry is restored. 
The elementary dislocation carrying the unit
of Burger vector is the `stripe coming to an end', Fig. 3b. However, this
is at the same time the topological excitation of the sublattice parity
order (`stripe dislocation'). Besides destroying the charge order,
it also  causes a seam
of sublattice parity mismatch extending to infinity. This injects frustration
in the spin system and as long as the  correlation length in the spin 
system is large compared to the lattice constant,  the system might want to
avoid this penalty by binding two such stripe dislocations into a single
`charge' dislocation carrying twice the Burger vector, Fig. 3a. If this
happens is a matter of microscopic numbers. However, when it happens 
the $Z_2$ local symmetry becomes alive.

\begin{figure}[t]
\hspace{0.00 \hsize}
\epsfxsize=0.9\hsize
\epsffile{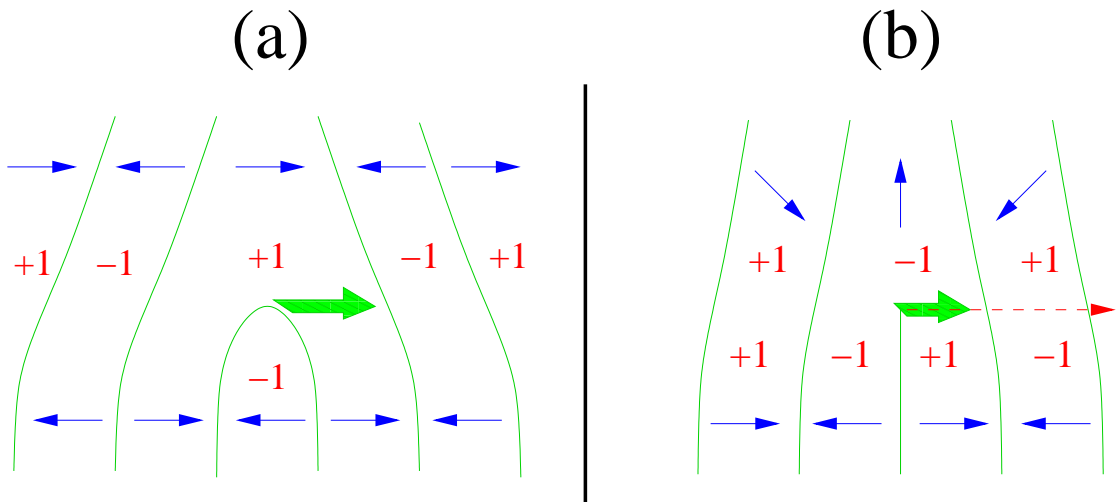}
\caption{Two types of topological defects can be distinguished in the stripe
phase: the charge- (a) and stripe (b) dislocations. The stripe dislocation
carries the minimal topological charge (Burger vector, thick arrow) relative
to the charge order. However, it is at the same time destroying the 
sublattice parity order, causing frustration in the spin system (thin arrows).
Although the charge dislocation carries a doubled Burger's vector, it
does not affect the sublattice parity order. When charge dislocations 
quantum-proliferate, sublattice parity turns into the Ising gauge field,
and the unbinding of charge dislocations into stripe dislocations corresponds
with the confinement phenomenon.} 
\label{fig3}\end{figure}

Let us first introduce the Ising gauge theory in its minimal form\cite{kogut}. 
Imagine a square lattice characterized
by Ising degrees of freedom $(1, 0), (0,1)$ defined on every bond $<ij>$
on the lattice, see fig. 4a. The Hamiltonian defining the $Z_2$ theory is,

\begin{equation}
H_{gauge}  =  -K \sum_{\Box} \sigma^3 \sigma^3 \sigma^3 \sigma^3
- \sum_{<ij>} \sigma^1_{ij}
\label{Z2gauge}.
\end{equation}

Here, $\sigma^{1,3}$ are just the Pauli-matrices residing on
bond variables, while
$\sum_{\Box} \sigma^3 \sigma^3 \sigma^3 \sigma^3$ is a short hand for: 
`pick the midpoint of a plaquette (dual lattice, $\Box$) and multiply the 
eigenvalue of the $\sigma^3$ operators of all 4 bonds surrounding 
this plaquette and sum the outcomes over the dual lattice.' The
plaquette (`flux') operator $\sigma^3 \sigma^3 \sigma^3 \sigma^3$ is also Ising
valued: it has eigenvalue $+1$ and $-1$ when the number of $+1$ bonds is even
or odd, respectively (Fig. 4). The {\em  local} symmetry
generator $P_i = \Pi_j \sigma^1_{ij}$ flips all bonds leaving an
arbitrary site $i$  and since this operation does not change the (un)evenness
probed by the flux operator while it surely commutes with the `kinetic' energy
$\sim \sum \sigma^1$ it commutes with the Hamiltonian. $P_i$ is the generator
of the Ising gauge transformations.

\begin{figure}[t]
\hspace{0.00 \hsize}
\epsfxsize=0.9\hsize
\epsffile{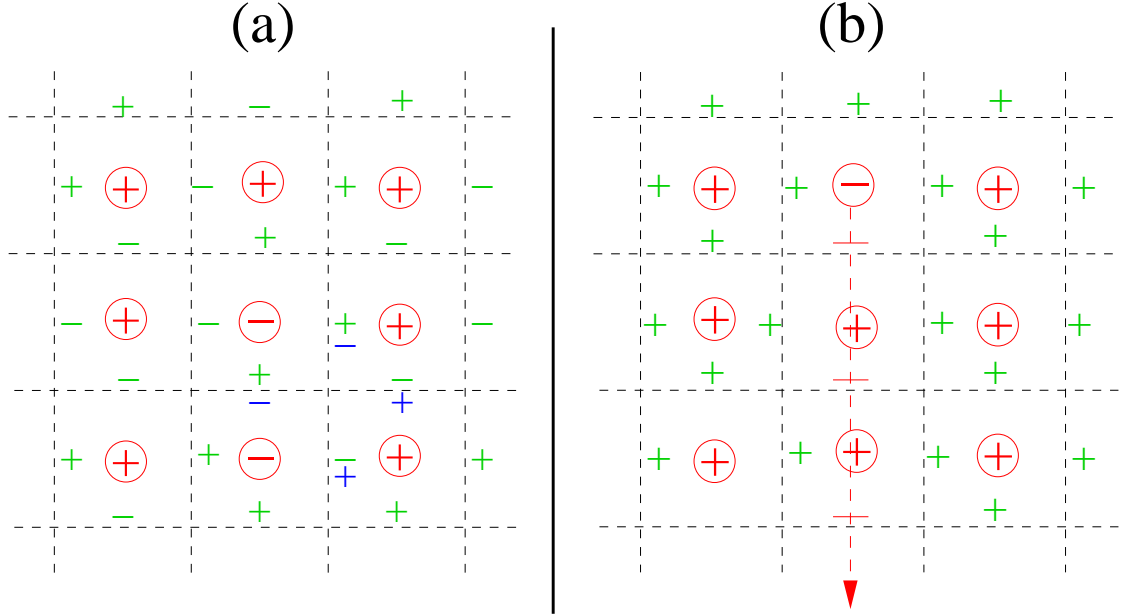}
\caption{The bond ($\pm$)- and flux ($\pm$ in the circles) variables
of the Ising gauge theory. A gauge transformation is indicated 
centered on the site shared by the lower-right four plaquettes 
in (a). It is left as an exercise to demonstrate that by repeated 
gauge transformations all bonds can be made positive 
(unitary gauge) except for the
bond shared by the two negative fluxes (bound vison pair). The
incipient order of the deconfining state is destroyed when visons,
isolated minus fluxes, proliferate and these carry half-infinite lines
of minus bonds as can be seen in unitary gauge (b).}  
\label{fig4}\end{figure}

The pure gauge theory Eq. (\ref{Z2gauge}) has two phases in 2+1D 
as function of $K$, separated by a continuous phase transition at
a critical coupling $K_c$, called the deconfining (large $K$) and
confining (small $K$) phases. How to understand these in terms of
the stripe physics? Imagine that every site on the gauge theory lattice
corresponds with a patch of size $\xi_c$ (charge correlation length) in
the real cuprate lattice. Interpret now the gauge bonds as being responsible
for the relative orientations of the sublattice parity on neighboring
patches: pick for instance $A - B - 0 - A - B$ on patch $i$ (`0' is a stripe)
to find $A - B - 0 - A - B$ on a neighboring patch $j$ when the bond
variable has value $+1$ and $B - A - 0 - B - A$ when the bond is $-1$.
Consider $K \rightarrow \infty$ so that all fluxes are $+1$ and consider
as representative gauge fix the unitary gauge where all bonds are positive.
This implies that all stripe patches have the same sense of sublattice
parity, like the ordered stripes. However, act once  with $P_i$
at site $i$: this will flip the sublattice parity at patch $i$
from $\cdots - A - B - \cdots$ into $\cdots - B - A - \cdots$ and
repeating these gauge transformations everywhere else one obtains 
a state characterized by the fact that sublattice parity is either
$\cdots - A - B - \cdots$ or $\cdots - B - A - \cdots$ although it
is impossible to say which choice is actually made. The reader should
recognize that this `deconfining' state is identical to 
a state characterized
by intact but delocalized domain walls: the charge fluctuations have
turned into the $Z_2$ gauge transformations $P_i$!

How to understand the confining state? The mathematical beauty of the gauge
theory lies in Wegner's discovery\cite{wegner} 
of an exact duality transformation. In
the Hamiltonian language\cite{kogut} this is easy to understand. 
Define the following operators
living on the sites $i^*$ of the dual (midpoints of plaquettes) lattice:
$\mu^1_{i^*} = \sigma^3 \sigma^3 \sigma^3 \sigma^3$  and
$\mu^3_{i^*} = \Pi_{n= - \infty}^{i^*} \sigma^1_{n}$. The operator 
$\mu^1$ just
measures the gauge flux but it is in $\mu^3$ that we encouter 
the novelty: start
at the middle of plaquette $i^*$ and draw a line to infinity of arbitrary
shape except that it crosses two bonds of every plaquette different from $i*$,
and flip the spin on every bond which is crossed by the line (Fig. 4b). 
It is easy to
check that $\mu^1$ and $\mu^3$ commute like Pauli matrices and using simple
operator identities the Hamiltonian becomes in terms of these operators,

\begin{equation}
H_{dual} = - K \sum_{i^*} \mu^1_{i^*} - \sum_{<i^*j^*>} \mu^3_{i^*} 
\mu^3_{j*}.
\label{Z2dual}
\end{equation}

This is just the simple global Ising model in 2+1D in the presence of
a transversal field! For $K \rightarrow 0$ the original gauge theory 
would be strongly disordered (confinement) and we discover that this
disordered state actually corresponds with simple Ising order in terms
of the topological excitations, the `visons', created by $\mu^3$ (Fig. 4b)! 
Local-global dualities of this kind are at the heart of Abelian 
gauge theories in 2+1D and it is believed that similar concepts are
behind the confinement in 3+1D quantum chromodynamics. In the present
context, it is particular significant that the confinement transition
is just a meat-and-potato 3D Ising universality class quantum
phase transition.

What is the meaning of confinement in the stripe interpretation? It is
simple: take the unitary gauge and let the gauge seam of minuses  
conseqeuently created by $\mu^3$ be along a straight line starting in the 
middle of the plain 
to disappear to $y \rightarrow - \infty$ (Fig. 4b). Remembering that these bond
values are about the relative orientation of sublattice parity this is
just identical to the stripe dislocation, Fig. 3b!  Hence,
the confining state corresponds with the stripe-fractionalized state and
its geometrical meaning is that space is neither $\cdots - A - B - \cdots$ 
nor $\cdots - B - A - \cdots$. Sublattice parity is destroyed by the
visons/stripe dislocations.

\section{Topological order and  cuprate criticality.}

Let us now turn to the cuprate superconductors. It is widely believed
that the regularities observed in the normal state of the optimally
doped superconductors reveals the presence of a zero temperature
quantum phase transition. It is most natural to expect that this 
has to do with a spontaneous symmetry breaking in the quantum vacuum,
implying some form of `order' is likely found on the underdoped side. However,
this order has not been seen experimentally and is therefore called
`hidden order'.

\begin{figure}[t]
\hspace{0.00 \hsize}
\epsfxsize=0.9\hsize
\epsffile{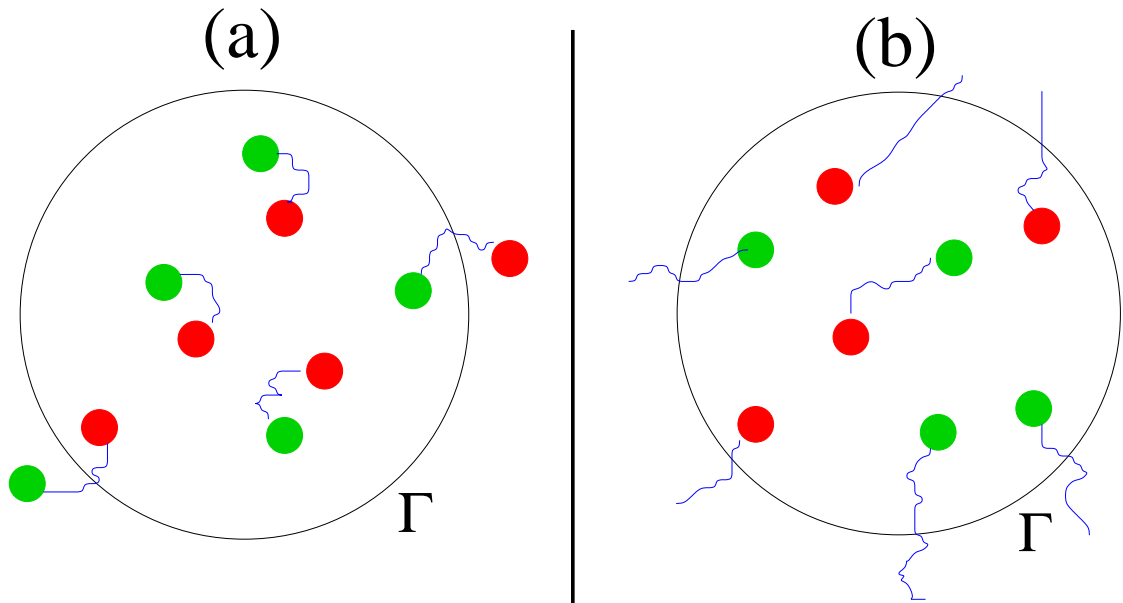}
\caption{The only device which can measure the stripe-topological order, 
and its disappearance at the confinement transition, is the non-local
correlation function well known in Gauge theory: the Wilson loop. In
the deconfining state, the loop is fluctuated by the gauge seams
connecting bound stripe dislocations, and these cause the Wilson loop
to decay with its parameter. In the confining state, the probability
of a seam crossing a loop is proportional to the enclosed area because
stripe dislocations occur freely.}
\label{fig5}\end{figure}

It is a fascinating possibility that this quantum phase transition
is about the confinement transition discussed in the above. The key is that
stripe-deconfinement order is a topological order which can only be
detected by truly topological means, and experimental physicists 
cannot build the machines to measure it. This order, and its destruction,
can only be directly measured by the Wilson loop. This measuring machine
looks as follows. First an operator $\tilde{\sigma}^3_i$ 
has to be designed measuring if at a particular site (or bond) in
the cuprate planes a domain wall is absent ($+1$) or present $(-1)$
(see references \cite{philmag},\cite{kruis}). 
Define a large closed contour on the plane $\Gamma$
and calculate the expectation value of the product of all  
$\tilde{\sigma}^3$ on the sites/bonds lying on the perimeter of the
contour,

\begin{equation}
O_{Wilson} (\Gamma) = \langle \Pi_{\Gamma} \tilde{\sigma}^3_{i \epsilon 
\Gamma} \rangle.
\label{Wilson}
\end{equation}

When the stripes would be perfectly connected, every stripe entering the loop
would have to come out again meaning that the number of stripes crossing the
loop would always be even, with the consequence that upon sending
the perimeter $l$ of the loop to infinity $O_{Wilson} \rightarrow 1$.
However, in reality, local (virtual) stripe unbindings will always occur 
(Fig. 5a).
A stripe dislocation-antidislocation pair might be created right
at the perimeter of the loop adding an unevenness fluctuation which
will cause the Wilson loop to decay exponentially. However, the 
probability for this to happen scales with the perimeter and accordingly
$O_{Wilson} \sim \exp (- \alpha l)$, the `perimeter law'\cite{kogut}. 
On the other hand,
in the confining state stripe end points occur  as real excitations and 
accordingly
unevenness fluctuations are now proportional to the number of stripe
end points occurring anywhere within the loop (fig. 5b). As a consequence,
$O_{Wilson} \sim \exp (- \alpha l^2)$, the famous area law.

The bad news is that the Wilson-loop machine seems beyond the capacities
of present day condensed matter experimentation and all what remains are
{\em indirect} ways to look for signatures of the transition -- in fact
the same problem is faced by the high energy physicists working on QCD 
confinement. One key question in this regard is how the gauge 
fields which become massless at the confinement transition communicate
with the fermionic degrees of freedom presumably also present at low
energies (nodal fermions). This is a difficult subject which is beyond
the scope of the present tutorial discussion and we refer to an interesting
recent discussion by Sachdev and Zhang\cite{sachdev2}.

Yet another way to look for circumstantial evidence is the 
subject of paper II\cite{strifractII}. Fully connected stripes protect 
the spin system
against frustrations. Although there are other sources of spin 
fluctuations, squeezed space is still bipartite as long as the
stripes are intact. On the other hand, a free vison/stripe dislocation
corresponds with a hard frustration in the spin system, destroying
colinear spin order globally (Fig. 3b). If the conditions are right, it is
imaginable that even when the stripes are delocalized the spin order
persists. In squeezed space this is just an antiferromagnet. However,
in full space the antiferromagnet  is fluctuated by the sublattice
parity flips, turning the staggered order parameter  in one which
is minus itself. This is the quantum spin-nematic order, which is at the
center of attention in paper II. As we will argue, also 
in the case of the spin-nematic it is a bit of a game of hide and seek, 
but now the experimental physicist has a fair chance to find the 
suspect.

Acknowledgments.  We acknowledge helpful discussions 
with S. Sachdev, E. Demler, S.A. Kivelson, H.V. Kruis, I. McCulloch,
T. Senthil and  M.P.A. Fisher. This work was supported by the Dutch
Science Foundation NWO/FOM.


\end{document}